\documentclass[aps, prb, twocolumn]{revtex4-2}

\usepackage{amsmath,amsfonts,amssymb}
\usepackage{graphicx}
\usepackage{hyperref}
\usepackage{epstopdf}
\usepackage{float}
\usepackage{multirow}
\usepackage{braket}
\usepackage{bm}
\usepackage{url}
\usepackage{dcolumn}
\usepackage{listings} 
\usepackage{xcolor}

\begin{document}

\title{A Probabilistic Imaginary Time Evolution Algorithm Based on Non-unitary Quantum Circuit}

\author{Hao-Nan Xie$^{1}$}
\email{xiehn19@mails.tsinghua.edu.cn}

\author{Shi-Jie Wei$^{2}$}
\email{weisj@baqis.ac.cn}

\author{Fan Yang$^{1}$}

\author{Zheng-An Wang$^{2}$}

\author{Chi-Tong Chen$^{3,4}$}

\author{Heng Fan$^{3,2}$}

\author{Gui-Lu Long$^{1,2}$}
\email{gllong@mail.tsinghua.edu.cn}

\affiliation{$^{1}$Department of Physics, Tsinghua University, Beijing 100084, China}
\affiliation{$^{2}$Beijing Academy of Quantum Information Sciences, Beijing 100193, China}
\affiliation{$^{3}$Institute of Physics, Chinese Academy of Sciences, Beijing 100190, China}
\affiliation{$^{4}$School of Physical Sciences, University of Chinese Academy of Sciences, Beijing 100190, China}

\date{\today}

\begin{abstract}
Imaginary time evolution is a powerful tool applied in quantum physics, while existing classical  algorithms for simulating imaginary time evolution suffer high computational complexity as the quantum systems become larger and more complex. In this work, we propose a probabilistic algorithm for implementing imaginary time evolution based on non-unitary quantum circuit. We demonstrate the feasibility of this method by  solving the ground state energy of several quantum many-body systems, including H$_2$, LiH molecules and the quantum Ising chain. Moreover, we perform experiments on superconducting and trapped ion cloud platforms respectively to find the ground state energy of H$_2$ and its most stable molecular structure. We also analyze the successful probability of the algorithm, which is a polynomial of the output error and introduce an approach to increase the success probability by rearranging the terms of Hamiltonian. 
\end{abstract}

\maketitle

\section{Introduction}

Imaginary time evolution (ITE), as a mathematical tool, has impacted many problems in quantum physics, such as solving the ground state of a Hamiltonian \cite{ITP, gHFT, QITP}, studying finite temperature properties \cite{Verst04, White09, SNS21} and the quantum simulation of non-Hermitian systems \cite{Kama22, Okuma22}. The concept of ITE can be understood by defining the imaginary time $\beta=-it$ and substituting it into the Schr\"{o}dinger's equation, $i\partial_t\ket{\Phi}=\mathcal{H}\ket{\Phi}$, where $\mathcal{H}$ is a Hermitian Hamiltonian, which gives us the imaginary-time Schr\"{o}dinger's equation:
\begin{equation}
\label{eq:itSE}
-\partial_\beta\ket{\Phi_\beta}=\mathcal{H}\ket{\Phi_\beta}.
\end{equation}
Given the initial state $\ket{\Phi_0}$, the solution of Eq. \ref{eq:itSE} is $\ket{\Phi_\beta}=Ae^{-\beta\mathcal{H}}\ket{\Phi_0}$, where the corresponding evolution operator $e^{-\beta\mathcal{H}}$ is non-unitary, and $A$ is the normalization constant. In classical simulations, one can directly calculate $e^{-\beta\mathcal{H}}$ and apply it to the initial state vector $\ket{\Phi_0}$, or employ some classical techniques such as quantum Monte Carlo \cite{MC_ITE} and tensor networks \cite{PEPS_ITE}. However, the dimension of the Hilbert space grows exponentially with the size of the quantum system, making the tasks intractable for classical computers \cite{Feyn82}.

Quantum computer is one of the promising tools for efficiently simulating quantum systems \cite{Benioff, Manin, Feyn82, Lyd96, Abr97, Kitaev95, Asp05, Babbush14, Nielson_Chuang}. For the real time simulation, the evolution operator $e^{-it\mathcal{H}}$ can be realized directly or be simply decomposed into a sequence of unitary quantum gates; but it is not the case for imaginary time simulation, where the evolution operator $e^{-\beta\mathcal{H}}$ is non-unitary. Therefore alternative methods are required. Recently, some hybrid quantum-classical algorithms for simulating ITE have been proposed. For example, variational quantum simulation methods \cite{VQS_ITE, VQS_general, RT_IT} utilize variational ansatz and simulate the evolution of quantum states with classical optimization of parameters; quantum imaginary time evolution (QITE) \cite{QITE, effi_QITE} finds a unitary operator to approach the ideal ITE in each evolution step. The main drawbacks of these methods include systematic error due to fixed parametrization, complexity of classical optimization \cite{BP, VQA_NP}, limitation of correlation length \cite{effi_QITE, QITP}, etc.

In 2004, Terashima and Ueda \cite{non_U} proposed a method to implement non-unitary quantum circuit by quantum measurement. By applying unitary operations in the extended Hilbert space, one can obtain the desired final state in a certain subspace of the auxiliary qubits. Non-unitary quantum circuit has been widely used in simulating non-Hermitian dynamics \cite{Okuma22, Wu19}, linear combination of unitary operators (LCU) \cite{LCU}, full quantum eigensolver (FQE) \cite{FQE} and other algorithms. Based on non-unitary circuit, Ref. \cite{Prob_gate} proposes a form of non-unitary gate which applies to two-qubit ITE process; Refs. \cite{QITP, Okuma22} show the form of the unitary operator acting on the total Hilbert space to implement ITE, and give the examples of circuits for two-qubit cases.

In this work, we propose a probabilistic imaginary time evolution (PITE) algorithm which utilizes non-unitary quantum circuit with one auxiliary qubit. In contrast to previous works, we explicitly illustrates the construction of the required quantum circuits using single- and double-qubit gates, which applies to any number of qubits and more generic Hamiltonians. We numerically apply the PITE algorithm to calculate the ground-state energy of several physical systems, and perform experiments on superconducting and trapped ion cloud platforms. We also give a detailed analysis about the computational complexity of this method.

This paper is organized as follows. In section \ref{sec:method} we give a description of the PITE method. Section \ref{sec:result} shows some experimental and numerical simulation results, and the analysis about the error and the successful probability. In Section \ref{sec:generalized}, we discuss the generalization of the PITE algorithm to the cases where Hamiltonian is not composed of Pauli terms, and introduce a method to increase the success probability.

\section{Method}
\label{sec:method}

An $n$-qubit Hamiltonian, $\mathcal{H}=\sum_{k=1}^{m} c_k h_k$, is composed of $m$ Pauli product terms, in which $c_k$ is a real coefficient and $h_k=\otimes_{j=1}^n\sigma_{\alpha_j}^j$, where $\sigma_{\alpha_j}^j$ is a Pauli matrix or the identity acting on the $j$-th qubit, with $\alpha_j\in\{0,x,y,z\}$ (here we use the notation $\sigma_0=I$). We assume that the Hamiltonian does not contain the identity term $I^{\otimes n}$ because it merely shift the spectrum of Hamiltonian.

Our goal is to implement the non-unitary operator $e^{-\beta\mathcal{H}}$ in quantum circuits. We first apply the Trotter decomposition \cite{Trotter, Chernoff}
\begin{equation}
\label{eq:Decomp_Trotter}
e^{-\beta\mathcal{H}}=\left(e^{-c_1 h_1 \Delta t}\dots e^{-c_m h_m \Delta t}\right)^{\beta/\Delta t}+\mathcal{O}(\Delta t).
\end{equation}
For a single Trotter step, we wish to obtain $\ket{\Phi'}=e^{-c_k h_k\Delta t}\ket{\Phi}$. Define $\widetilde{T}_k=e^{-c_k h_k\Delta t}$. Due to the fact that $c_k h_k$ only has two different eigenvalues $\pm|c_k|$, each with the degeneracy of $2^{n-1}$, there exists a unitary $U_k$ satisfying
\begin{equation}
\label{eq:UchU}
U_k c_k h_k U_k^\dagger=-|c_k|\sigma_z^{l_k},\quad l_k\in\{1,\dots,n\}
\end{equation}
which is a single qubit operator. Thus we have
\begin{equation}
\label{eq:Tk}
\widetilde{T}_k=U_k^\dagger \exp\left(-|c_k|\sigma_z^{l_k}\Delta t\right)U_k.
\end{equation}

\begin{figure}[htbp]
\includegraphics[width=8cm]{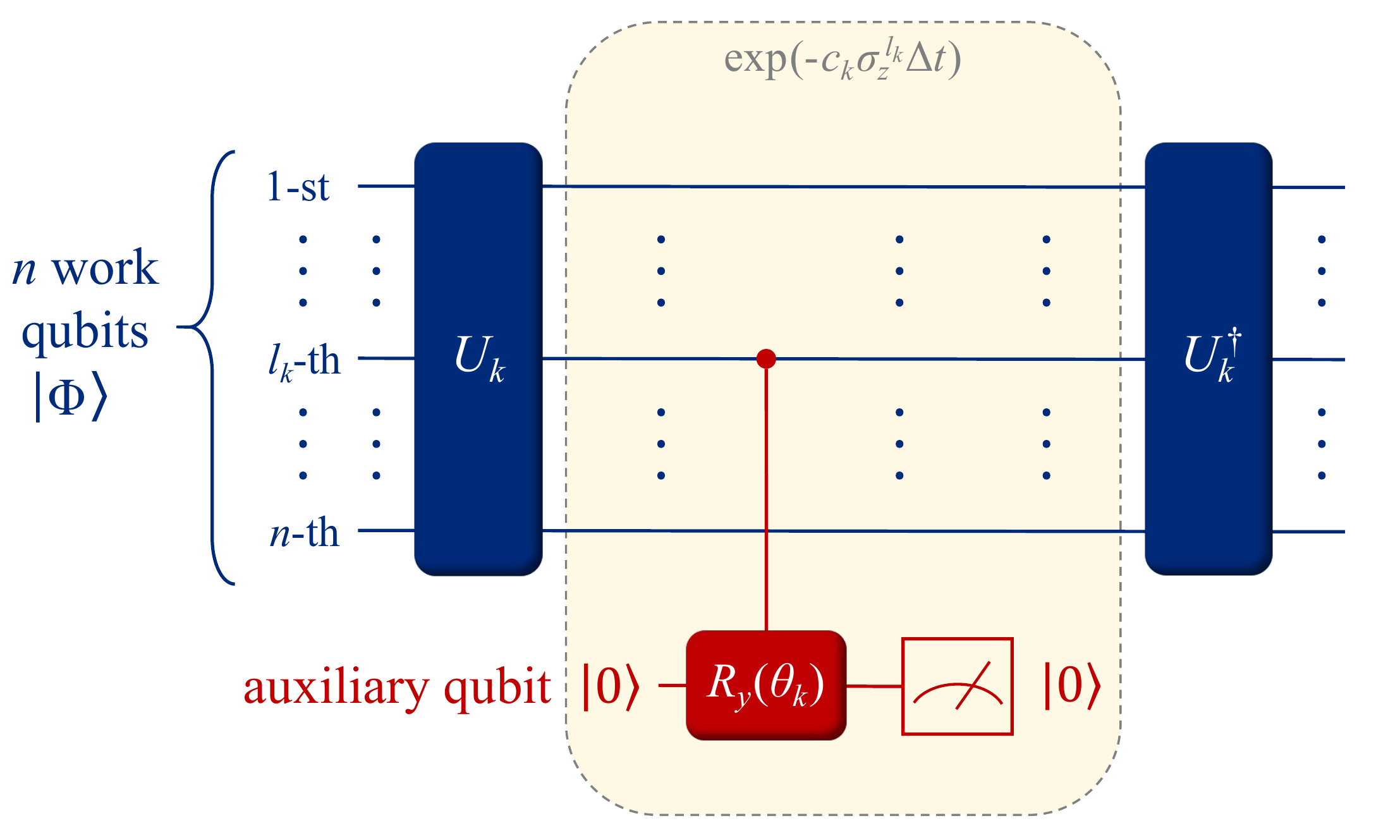}
\caption{Quantum circuit for implementing $\widetilde{T}_k$.}
\label{circuit:Tk}
\end{figure}

In Fig. \ref{circuit:Tk}, we show how to implement $\widetilde{T}_k$ in a quantum circuit. The construction of $U_k$ requires $\mathcal{O}(n)$ CNOT gates and single qubit gates at most (see Appendix \ref{app:Uk} for details). After the action of $U_k$, we write the work-qubit state as $U_k\ket{\Phi}=a_0\ket{\psi_0}+a_1\ket{\psi_1}$, where $\ket{\psi_0}$ and $\ket{\psi_1}$ are the projection of $U_k\ket{\Phi}$ on the subspace where the $l_k$-th qubit is $\ket{0}$ and $\ket{1}$, respectively. In Appendix \ref{app:Uk} we will show that $a_0,a_1$ are also the amplitude of the projection of $\ket{\Phi}$ on the ground-state subspace and excited-state subspace of $c_k h_k$, respectively.

To realize $e^{-|c_k|\sigma_z\Delta t}$ on the $l_k$-th work qubit, we add an ancillary qubit $\ket{0}$, and apply the controlled-$R_y$ operation $\ket{0}\bra{0}\otimes I+\ket{1}\bra{1}\otimes R_y(\theta_k)$ on the $l_k$-th work qubit and the ancilla qubit, where
\begin{equation}
\label{eq:Ry}
\begin{split}
&R_y(\theta)=e^{-i\theta\sigma_y/2}=\begin{pmatrix} \cos{\theta/2} & -\sin{\theta/2} \\ \sin{\theta/2} & \cos{\theta/2} \end{pmatrix}, \\
&\theta_k=2\cos^{-1}{(e^{-2|c_k|\Delta t})},
\end{split}
\end{equation}
which gives the state
\begin{equation}
\label{eq:after_Ry}
\begin{split}
&a_0\ket{\psi_0}\ket{0}_\text{anc}+a_1 e^{-2|c_k|\Delta t}\ket{\psi_1}\ket{0}_\text{anc} \\
&+a_1\sqrt{1-e^{-4|c_k|\Delta t}}\ket{\psi_1}\ket{1}_\text{anc}.
\end{split}
\end{equation}
Then we measure the ancilla qubit, and if the result is 0, we obtain
\begin{equation}
\label{eq:after_measure}
\sqrt{\frac{1}{|a_0|^2+|a_1|^2 e^{-4|c_k|\Delta t}}}\left(a_0\ket{\psi_0}+a_1 e^{-2|c_k|\Delta t}\ket{\psi_1}\right)
\end{equation}
which is equivalent to the result of $e^{-|c_k|\sigma_z\Delta t}$ acting on the $l_k$-th work qubit up to normalization.

The probability of obtaining 0 in the measurement of ancilla is $|a_0|^2+|a_1|^2 e^{-4|c_k|\Delta t}$. If we success, then the last step is to apply $U_k^\dagger$ on the work qubits. The output state will be exactly $\widetilde{T}_k\ket{\Phi}$ up to normalization.

In summary, the non-unitary operator $\widetilde{T}_k$ acting on a quantum state $\ket{\Phi}$ can be written as
\begin{equation}
\label{eq:Tk:comp}
\widetilde{T}_k\ket{\Phi}=U_k^\dagger\left[\bra{0}_\text{anc}\mathcal{R}_k\cdot\left(U_k\ket{\Phi}\otimes\ket{0}_\text{anc}\right)\right]
\end{equation}
up to a constant coefficient, where $U_k$ transforms $h_k$ into a Pauli matrix acting on the $l_k$-th work qubit, and $\mathcal{R}_k$ represents the controlled-$R_y$ gate acting on the $l_k$-th work qubit and the ancilla qubit, with the rotation angle $\theta_k=2\cos^{-1}\left(e^{-2|c_k|\Delta t}\right)$.

\section{Results}
\label{sec:result}

\subsection{Calculation of H$_2$, LiH and quantum Ising model}

To illustrate the performance of the PITE algorithm, we apply the algorithm in the calculation of the ground-state energy of three physical systems: H$_2$ molecules, LiH molecules, and a quantum Ising spin chain with both transverse and longitudinal fields. The calculations of H$_2$ are carried out on the Quafu's 10-qubit superconducting quantum processor and IonQ's 10-qubit trapped ion QPU, and the calculations of LiH and the Ising model are carried out on a numerical simulator to study the influence of noises and the success probability of the algorithm.

To calculate the ground state of H$_2$ and LiH on a quantum computer, we first need to encode the molecular Hamiltonian onto qubits. Here we choose the STO-3G basis \cite{STO3G} and use the Jordan-Wigner transformation (JWT) (see details in Appendix \ref{app:molecules}). We eventually obtain Hamiltonians composed of Pauli matrices $\mathcal{H}(R)=\sum_k c_k(R)\sigma_{\alpha_1}^1\dots\sigma_{\alpha_n}^n$, which can be acted on $n$ qubits, and the coefficients $c_k$ vary with the interatomic distance $R$. In this way the Hamiltonian for H$_2$ and LiH can be encoded onto 4 and 6 qubits, respectively. Further mapping is applied on H$_2$ to compactly encode the H$_2$ Hamiltonian onto 2 qubits (see details in Ref. \cite{Colless18}), which gives the H$_2$ Hamiltonian in the form of
\begin{equation}
\label{eq:H2}
\begin{split}
\mathcal{H}_{\text{H}_2}&=c_0(R)+c_1(R)\sigma_z^1+c_1(R)\sigma_z^2 \\
&+c_2(R)\sigma_z^1\sigma_z^2+c_3(R)\sigma_x^1\sigma_x^2,
\end{split}
\end{equation}
where the coefficients at different $R$ are available in Appendix \ref{app:molecules}. The Hamiltonian for LiH at its lowest-energy interatomic distance (bond distance) is given explicitly also in Appendix \ref{app:molecules}.

\begin{figure}[htbp]
\includegraphics[width=8.5cm]{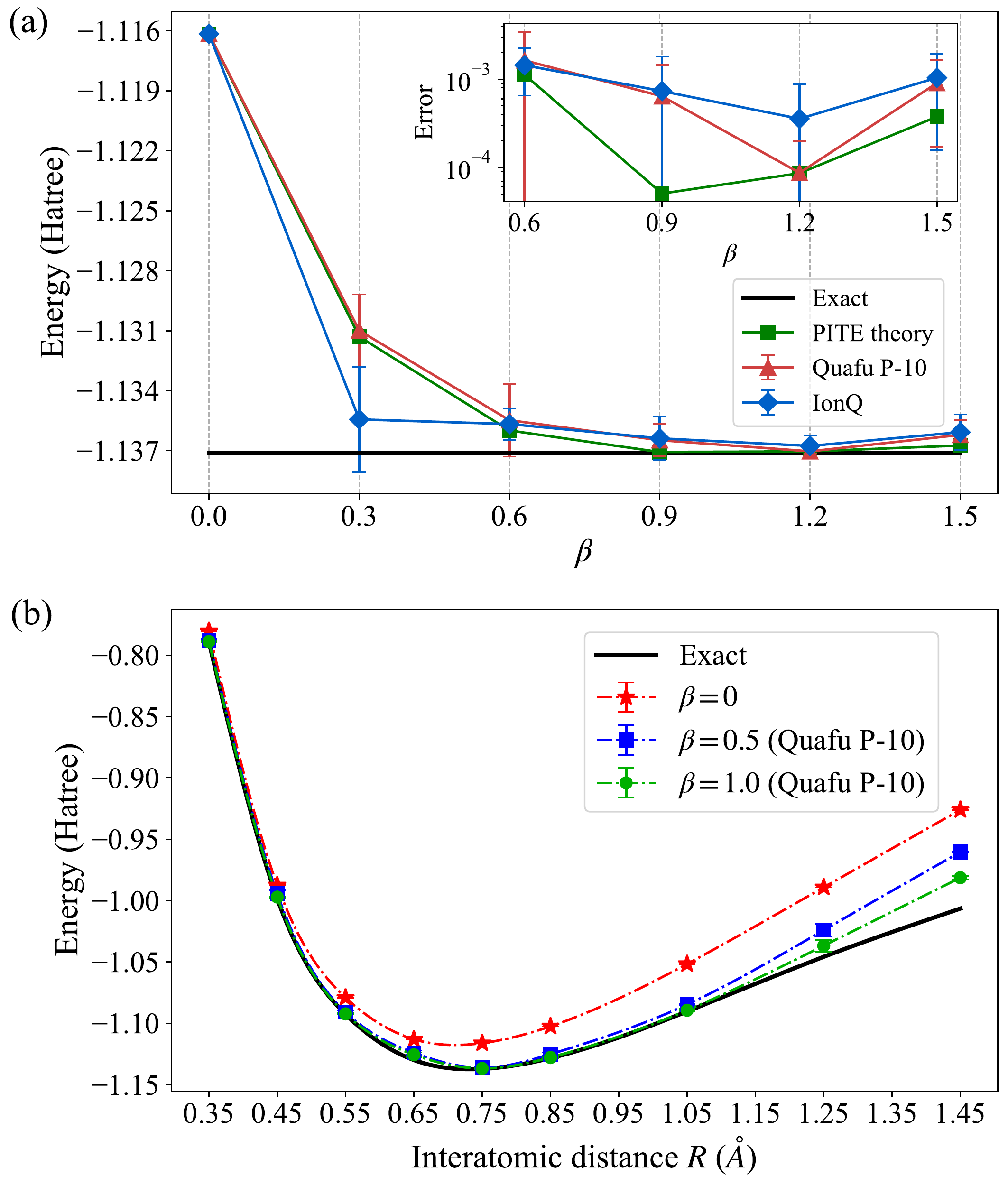}
\caption{Experimental results of PITE on Quafu and IonQ cloud platform. (a) H$_2$ energy expectation value as a function of $\beta$, at a fixed interatomic distance $R=0.75$\AA. The identity term in the Hamiltonian is considered when calculating the energy value, but not considered when executing the algorithm. (b) H$_2$ energy as a function of $\beta$ and the interatomic distance $R$. The black line is the exact ground state energy obtained by diagonalization.}
\label{expr:eR}
\end{figure}

In our experiments, we use 2 qubits as the work qubits to represent the H$_2$ molecule, and 1 qubit as the ancilla qubit. The Hatree-Fock state of H$_2$ is $\ket{\Phi_\text{HF}}=\ket{00}$ in the qubit representation, which is chosen as the initial state of the work qubits. Following the PITE method, we first do the calculation at a fixed interatomic distance $R=0.75$\AA, and the experiments are carried out on Quafu's superconducting QPU P-10 and IonQ's trapped ion QPU (see information about Quafu in Appendix \ref{app:Quafu}). After each Trotter step, the quantum state of work qubits is tomographed, with 2000 shots on Quafu and 1000 shots on IonQ, and then we use the state to calculate the energy value, and set it as the initial state for the next step. The quantum circuits and related details are shown in Appendix \ref{app:exp}. The results of the energy expectation value as a function of the imaginary time $\beta$ are shown in Fig. \ref{expr:eR}(a), compared with the theoretical PITE results. As $\beta$ increases, the energy rapidly converges to the exact solution in 5 evolution steps, within an error of $\sim10^{-4}$ a.u, which is in the chemical precision.

To obtain the most stable molecular structure, we vary the interatomic distance and plot the potential-energy surface for H$_2$ molecule, as shown in Fig. \ref{expr:eR}(b). The series of experiments is only carried out on Quafu's P-10, and the results ($\beta=0.5$ and $\beta=1$) are compared with the Hatree-Fock state energies ($\beta=0$) and the exact ground-state energies obtained by diagonalization. The lowest energy in the potential-energy surface corresponds to the bound distance of H$_2$ molecule, which is around 0.75\AA.

\begin{figure*}[htbp]
\includegraphics[width=16cm]{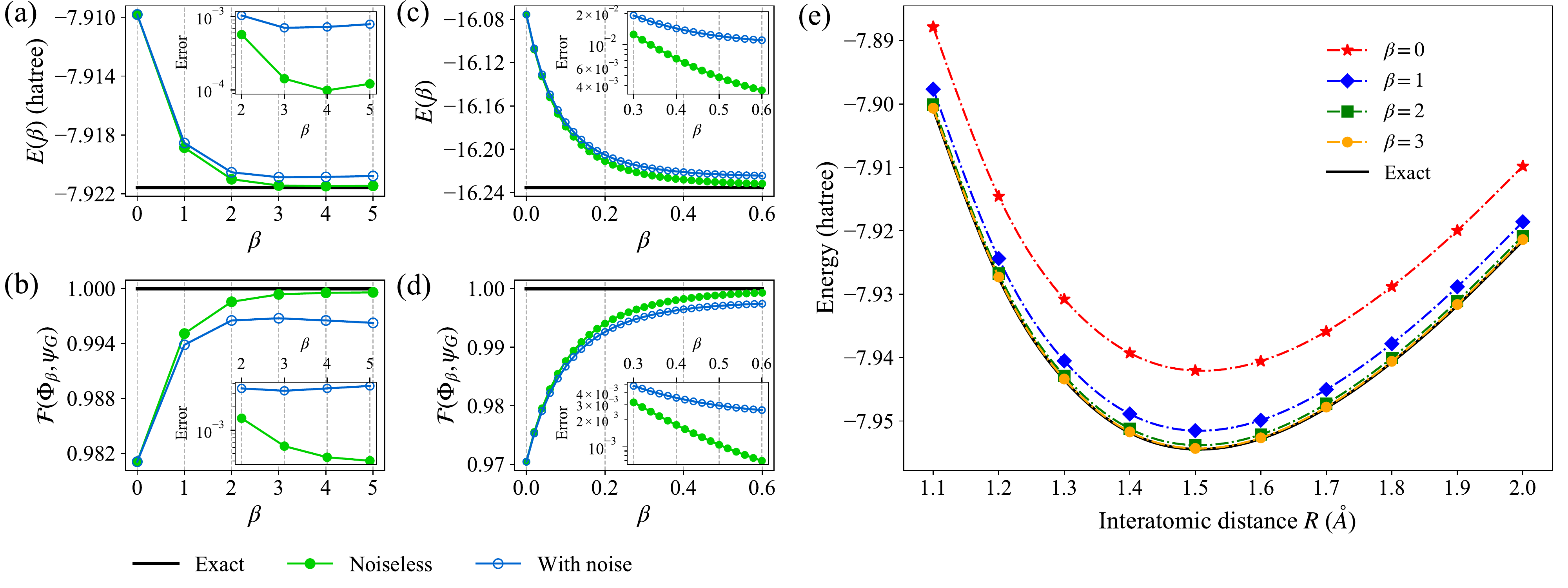}
\caption{Simulation results of PITE. For LiH, the energy (a) and fidelity $\mathcal{F}(\ket{\Phi_\beta},\ket{\psi_G})$ (b) as a function of $\beta$, respectively, at a fixed interatomic distance $R=2.0$\AA. The identity term in the Hamiltonian is considered when calculating the energy value, but not considered when executing the algorithm. For quantum Ising chain, the energy (c) and fidelity $\mathcal{F}(\ket{\Phi_\beta},\ket{\psi_G})$ (d) as a function of $\beta$. Parameters are chosen as $n=10, J=1, g=1.2, h=0.3$. In the noise simulation of both cases, the relaxation and dephasing parameters are set as $\epsilon_\text{r}=\epsilon_\text{d}=1\text{e-5}$. (e) LiH energy as a function of $\beta$ and the interatomic distance. The black line is the exact ground state energy obtained by diagonalization.}
\label{numr:efR}
\end{figure*}

In our numerical simulations, we calculate the ground-state energy of LiH and quantum Ising chain to study the success probability and the influence of noises. For LiH, we use 6 work qubits and 1 ancilla qubit. The Hatree-Fock state is $\ket{\Phi_\text{HF}}=\ket{110000}$ in the qubit representation. Here $\ket{\Phi_\text{HF}}$ is very close to the exact ground state, so it takes few steps for the state to converge. Therefore, to show more about the convergence process, we use a superposition of $\ket{\Phi_\text{HF}}$ and an excited state, $\ket{\Phi_0}=\sqrt{0.99}\ket{\Phi_\text{HF}}+0.1\ket{000011}$, as the initial state. Fig. \ref{numr:efR}(a) and Fig. \ref{numr:efR} (b) show the convergence of the energy $E(\beta)$ and the fidelity $\mathcal{F}$ as a function of $\beta$, respectively, where $\mathcal{F}$ is the fidelity between $\ket{\Phi_\beta}$ and the exact ground state $\ket{\psi_G}$. The influence of quantum noises are studied by applying quantum channels on all qubits before each measurement of the ancilla qubit. The noise is described by
\begin{equation}
\label{eq:channel}
\mathcal{E}(\rho)=\sum_{\nu=1}^3 \hat{E}_\nu \rho\hat{E}_\nu^\dagger
\end{equation}
with Kraus operators
\begin{equation}
\label{eq:Kraus}
\begin{split}
&\hat{E}_1=\begin{pmatrix} 1 & 0 \\ 0 & \sqrt{1-\epsilon_\text{r}-\epsilon_\text{d}} \end{pmatrix}, \\
&\hat{E}_2=\begin{pmatrix} 0 & \sqrt{\epsilon_\text{d}} \\ 0 & 0 \end{pmatrix}, \quad
\hat{E}_3=\begin{pmatrix} 0 & 0 \\ 0 & \sqrt{\epsilon_\text{r}} \end{pmatrix}
\end{split}
\end{equation}
where $\epsilon_\text{r}$ and $\epsilon_\text{d}$ are relaxation parameter and dephasing parameter, respectively (see details in Appendix \ref{app:noise}). The simulation results are also shown in Fig. \ref{numr:efR}(a) and Fig. \ref{numr:efR}(b), which indicates the energy still converges to the exact solution within an error of $\sim10^{-3}$ a.u. under the noisy condition.

We vary the interatomic distance and plot the potential-energy surface for LiH molecule, as shown in Fig. \ref{numr:efR}(e). The simulation results at different $\beta$ are compared together, as well as the exact solution obtained by diagonalization. The lowest energy in the potential-energy surface corresponds to the bound distance of LiH molecule, which is around 1.5\AA.

Finally, we compute the ground-state energy of a $n$-site cyclic quantum Ising chain with Hamiltonian
\begin{equation}
\label{eq:Ising}
\mathcal{H}=-J\sum_{j=1}^n\left(\sigma_z^j\sigma_z^{j+1}+g\sigma_x^j+h\sigma_z^j\right),
\ \ \sigma_z^{n+1}=\sigma_z^1
\end{equation}
where $g$ and $h$ are the magnitudes of the transverse and longitudinal fields, respectively. In the simulation, the state of the work qubits is initialized as $\ket{\Phi_0}=\left(\cos{\frac{\phi_0}{2}}\ket{0}+\sin{\frac{\phi_0}{2}}\ket{1}\right)^{\otimes n}$, where $\phi_0$ is chosen to minimize the initial energy $E(\beta=0)=\braket{\Phi_0|\mathcal{H}|\Phi_0}$. In FIG. \ref{numr:efR}(c) and FIG. \ref{numr:efR}(d), we show the energy and fidelity obtained by PITE algorithm, as well as the influence of noises. The results show that the energy converges to the exact value within an error of $\sim10^{-3}$ in the noiseless case and $\sim10^{-2}$ in the noisy case.

\subsection{Analysis of computational complexity}
\label{sub:analysis}

Here, we analyze the computational complexity of the PITE algorithm in three aspects: gate complexity, the number of evolution steps, and measurement complexity due to the probabilistic measurements. For gate complexity, the operator $e^{-c_k h_k\Delta t}$ can be decomposed into $\mathcal{O}(n)$ basic gates. Therefore, we need $\mathcal{O}(nmL)$ basic gates for an $m$-term Hamiltonian in total, where $L=\beta/\Delta t$ is the number of evolution steps.

Next we need to determine $L$. Equivalently, we need to know $\beta$ if $\Delta t$ is given. It is essential to do this before the algorithm, because we do not want to measure the expectation values during the algorithm procedure, which will destroy the state of the work qubits and stop the algorithm. To determine $\beta$, we show after the evolution of imaginary time $\beta$, the fidelity between $\ket{\Phi_\beta}$ and exact ground state $\ket{\psi_G}$ is limited to
\begin{equation}
\label{eq:fidelity_beta}
\mathcal{F}(\Phi_\beta,\psi_G)\geq\frac{s_0}{s_0+(1-s_0)e^{-2\beta\Omega_1}}
\end{equation}
where $s_0$ is the initial fidelity at $\beta=0$, $\Omega_1$ is the gap between the first excited state and the ground state (see Appendix \ref{app:complexity} for proof). When the fidelity of the output state is greater than $1-\epsilon$, the imaginary time length
\begin{equation}
\label{eq:beta_error}
\beta=\mathcal{O}\left(\frac{1}{\Omega_1}\log\frac{1}{\epsilon}\right)
\end{equation}
is linearly dependent on the inverse of the energy gap of Hamiltonian, and logarithmically dependent on the inverse of output error. Here we haven't considered the error caused by Trotter decomposition (Eq. \ref{eq:Decomp_Trotter}), which could reduce the fidelity in Eq. \ref{eq:fidelity_beta} by $\mathcal{O}(\Delta t)$. This term could not be decreased by increasing $\beta$, but we can use the higher order decomposition \cite{Sorn99}
\begin{equation}
\label{eq:Decomp_2order}
\begin{split}
e^{-\beta\mathcal{H}}&=\left[\left(e^{-c_1 h_1 \Delta t/2}\dots e^{-c_M h_M \Delta t/2}\right)\right. \\
&\qquad\left.\times\left(e^{-c_M h_M \Delta t/2}\dots e^{-c_1 h_1 \Delta t/2}\right)\right]^{\beta/\Delta t}+\mathcal{O}(\Delta t^2)
\end{split}
\end{equation}
to reduce the error term to $\mathcal{O}(\Delta t^2)$.

We also wish to know the success probability of measurement. For the quantum circuit which implements $e^{-c_k h_k\Delta t}$, the probability of obtaining $\ket{0}$ from the ancilla is
\begin{equation}
\label{eq:prob_k}
P_k=|a_0|^2+|a_1|^2 e^{-4|c_k|\Delta t}
\end{equation}
where $|a_0|^2+|a_1|^2=1$. Take the lower bound of Eq. \ref{eq:prob_k} as $P_k\geq e^{-4|c_k|\Delta t}$. Then the total success probability after $\Delta t$ is limited to
\begin{equation}
\label{eq:prob_dt}
P(\Delta t)=\prod_{k=1}^{m}P_k\geq\exp\left(-4\Delta t\sum_k|c_k|\right).
\end{equation}
Thus we can give a rigorous lower bound (RLB) of the final success probability after imaginary time $\beta$ as
\begin{equation}
\label{eq:prob_f_rlb}
P_\text{final}\geq P_\text{RLB}=\exp\left(-4\beta\sum_k|c_k|\right)
\end{equation}
which is exponential to $\beta$ and the sum of $|c_k|$'s. Note that the RLB is reached when and only when $a_0=0$ in Eq. \ref{eq:prob_k} for all Pauli terms during the whole evolution process, which is the worst case and almost never occurs. In most cases, much greater success probability than the RLB can be reached (we'll show the results later). A more practical lower bound (approximate lower bound, ALB) for estimating success probability is given by
\begin{equation}
\label{eq:prob_f_alb}
\begin{split}
&P_\text{final}\geq P_\text{ALB} \\
&\sim\exp\left[-2\beta\left(E_G+\sum_k|c_k|\right)-\frac{(1-s_0)\Omega_\text{max}}{s_0\Omega_1}\left(1-e^{-2\beta\Omega_1}\right)\right].
\end{split}
\end{equation}
where $\Omega_\text{max}$ is the energy gap between the highest excited state and the ground state (see Appendix \ref{app:complexity}). Fig. \ref{numr:pr}(a) and Fig. \ref{numr:pr}(c) shows the success probability as a function of $\beta$ in the simulations of LiH and the Ising model, which indicates an exponential decay of success probability as $\beta$ increases. We can clearly see that ALB is a much better approximation to the simulation results than RLB in both cases. The results also indicate that the success probability is hardly affected by the noises.

Usually we merely care about the relation between measurement complexity and the output error. According to Eq. \ref{eq:fidelity_beta}, \ref{eq:prob_f_rlb} and \ref{eq:prob_f_alb}, the RLB and ALB of the final success probability when the output fidelity is greater than $1-\epsilon$ is
\begin{equation}
\label{eq:prob_error}
\begin{split}
&P_\text{RLB}=\mathcal{O}\left(\left(\frac{\epsilon}{1-\epsilon}\right)^{\kappa_0}\right) \\
&P_\text{ALB}=\mathcal{O}\left(\left(\frac{\epsilon}{1-\epsilon}\right)^{\kappa_1}\right)
\end{split}
\end{equation}
where $\kappa_0=2\sum_k |c_k|/\Omega_1$ and $\kappa_1=\left(E_G+\sum_k |c_k|\right)/\Omega_1$ depend on the spectrum of the Hamiltonian, but independent on the scale of the Hamiltonian. In Fig. \ref{numr:pr}(b) and Fig. \ref{numr:pr}(d), we show the success probability as a function of $\epsilon$ in our simulations. The success probability could be lower than the ALB when qubits are affected by the noises.

\begin{figure*}
\includegraphics[width=16cm]{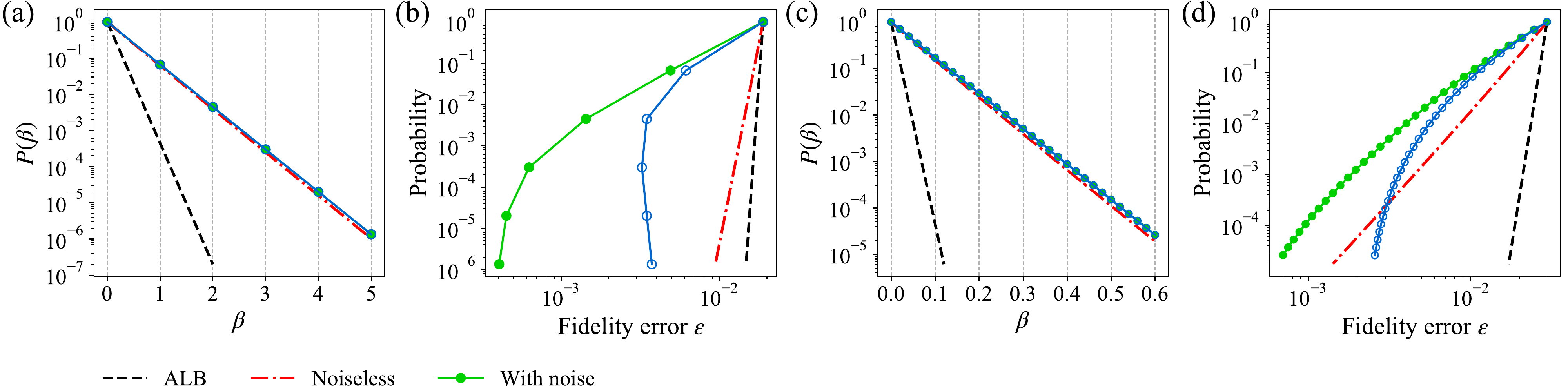}
\caption{Success probability of PITE in numerical simulations of LiH (a)(b) and quantum Ising chain (c)(d), as a funtion of $\beta$ (a)(c) and as a funtion of fidelity error $\epsilon$ (b)(d). For LiH, the RLB and ALB are calculated with dropping the identity term from the Hamiltonian.}
\label{numr:pr}
\end{figure*}

\section{Generalization of PITE algorithm}
\label{sec:generalized}

In this part we will generalize the PITE algorithm, for it's not necessary for the Hamiltonian to be a sum of Pauli terms. In general, the Hamiltonian is written as $\mathcal{H}=\sum_{k=1}^m H[k]$, and the ITE operator is decomposed as
\begin{equation}
\label{eq:Trotter_generalize}
e^{-\beta\mathcal{H}}=\left(e^{-H[1]\Delta t}\dots e^{-H[m]\Delta t}\right)^{\beta/\Delta t}+\mathcal{O}(\Delta t).
\end{equation}

Assume that the eigenvalues of each term $H[k]$ are easily known through classical calculation, and the corresponding eigenstates can be easily prepared on quantum computers. This assumption is true for most cases (such as $H[k]$'s are local terms). Denote $\lambda[k]_i$ as the eigenvalues of $H[k]$, and $\ket{\lambda[k]_i}$ as the corresponding eigenstates. The procedure of implementing the operator $e^{-H[k]\Delta t}$ on a quantum computer is described as following:
\begin{itemize}
\item[1. ] Define $\omega[k]_i=\lambda[k]_i-\lambda[k]_0$, where $\lambda[k]_0$ is the lowest eigenvalue of $H[k]$;
\item[2. ] Apply a unitary $U[k]$ to the work qubits, which transforms $\ket{\lambda[k]_i}$ into computational basis $\ket{x[k]_i}$;
\item[3. ] Add an ancilla qubit which is initialized as $\ket{0}$;
\item[4. ] Apply the gate $\sum_i \ket{x[k]_i}\bra{x[k]_i}\otimes R_y(\theta[k]_i)$, where $\theta[k]_i=2\cos^{-1}(e^{-\omega[k]_i \Delta t})$;
\item[5. ] Measure the ancilla qubit; If obtaining 0, continue the procedure; Else, start from beginning;
\item[6. ] Apply $U[k]^\dagger$ to the work qubits. End.
\end{itemize}
Note: it should be easy to implement $U[k]$ and $U[k]^\dagger$, because we have assumed $\ket{\lambda[k]_i}$ can be easily prepared on quantum computers.

Obviously, when $H[k]$'s are Pauli terms, this procedure degenerates into the original version of PITE algorithm. So it is a generalized PITE algorithm. Moreover, the generalized PITE can help us increase the success probability. We can prove the ALB of the final success probability of the generalized PITE is approximated by
\begin{equation}
\label{eq:prob_f_alb_gnrl}
\begin{split}
&P_\text{final}\geq P_\text{ALB} \\
&\sim\exp\left[-2\beta\left(E_0-\sum_k\lambda[k]_0\right)
-\frac{(1-s_0)\Omega_\text{max}}{s_0\Omega_1}\left(1-e^{-2\beta\Omega_1}\right)\right].
\end{split}
\end{equation}
Especially, when $H[k]=c_k h_k$ are Pauli terms, $\lambda[k]_0=-|c_k|$, which turns Eq. \ref{eq:prob_f_alb_gnrl} into Eq. \ref{eq:prob_f_alb}. We can see that when $\mathcal{H}$ is given, the only changeable part in Eq. \ref{eq:prob_f_alb_gnrl} is $\sum_k\lambda[k]_0$. We can divide $\mathcal{H}$ into $H[k]$'s in different ways, which enables us to increase $\sum_k\lambda[k]_0$ and enlargement the success probability.

\begin{figure}
\includegraphics[width=8.5cm]{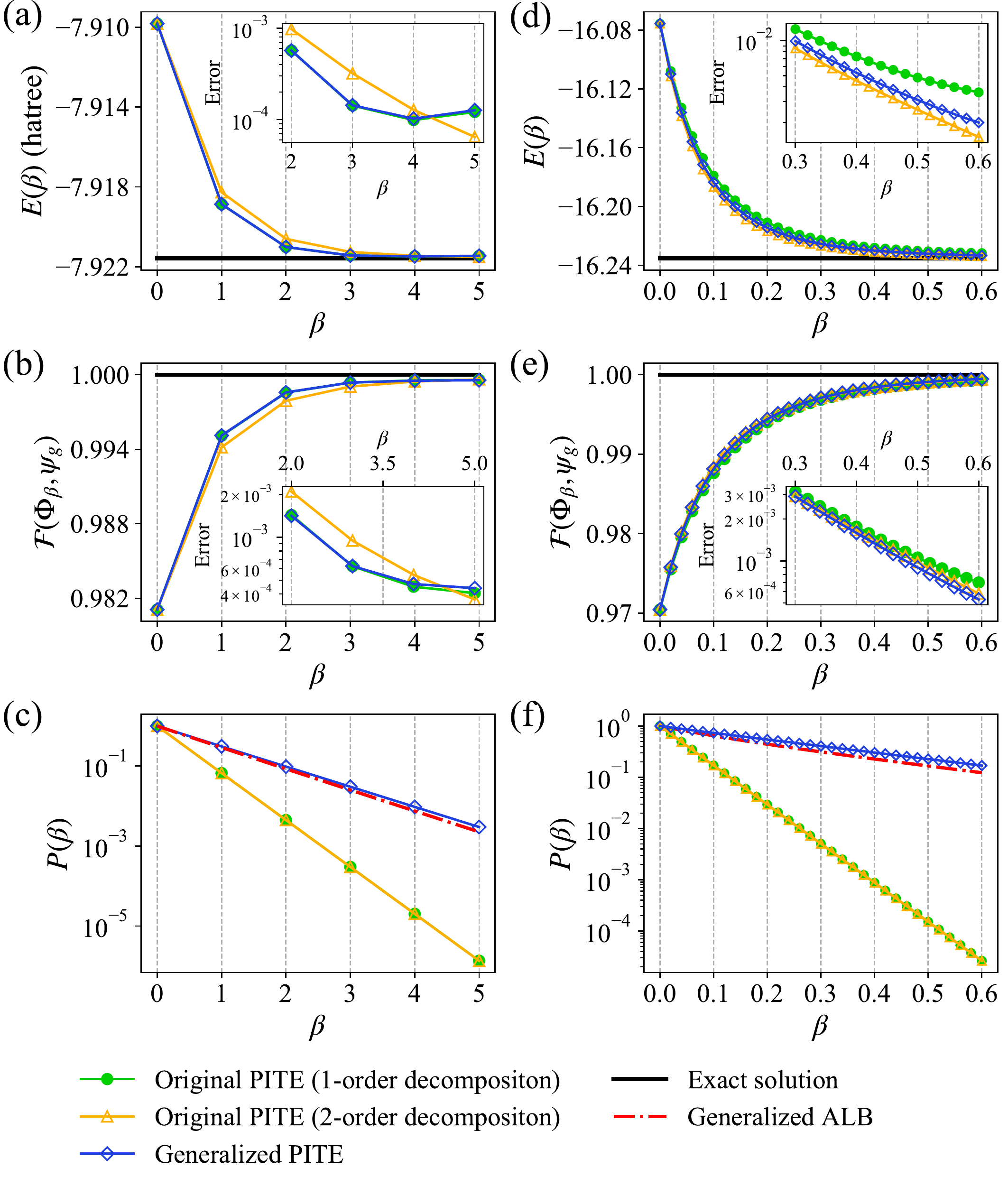}
\caption{Simulation results of generalized PITE. (a)(b)(c) LiH molecules. (d)(e)(f) Quantum Ising chain. Energy (a)(d), fidelity (b)(e), success probability (c)(f) as a function of $\beta$. For LiH, the generalized ALB is calculated with dropping the identity term from the Hamiltonian.}
\label{numr:gnrl}
\end{figure}

We apply the generalized PITE on the simulations of LiH (at $R=2.0$\AA) and quantum Ising model. Instead of taking $c_k h_k$ as $H[k]$, we rearrange these terms and divide the Hamiltonian in another way (see Appendix \ref{app:generalized} for details). The results are shown in Fig. \ref{numr:gnrl}, where the success probabilities obtained by generalized PITE are compared with that given by Eq. \ref{eq:prob_f_alb_gnrl}. The results indicate the generalized PITE has little effects on reducing the error, but performs much better on the success probability, which makes the PITE algorithm much more practical.

We also run the simulation using the original PITE with second-order decomposition (Eq. \ref{eq:Decomp_2order}). As shown in Fig. \ref{numr:gnrl}, the results indicate little improvement in the output error and the success probability from the previous results.

\section{Conclusions}
\label{sec:conclusion}

In this paper, we proposed a probabilistic algorithm for implementing imaginary time evolution, PITE, based on non-unitary quantum circuit, and show the explicit construction of the circuit which applies to any number of qubits. This algorithm can be applied in solving the ground state of a Hamiltonian. For an $n$-qubit Hamiltonian composed of $m$ Pauli terms, the algorithm returns the ground state within an error of $\mathcal{O}(\Delta t)=\mathcal{O}(\beta/L)$ using $\mathcal{O}(nmL)$ gates, with the success probability described in Eq. \ref{eq:prob_f_alb}. We demonstrated the feasibility and performance of this method with the example of H$_2$, LiH molecules and the quantum Ising chain model, in experiments and numerical simulations. We also generalized this approach to the cases where the Hamiltonian is not composed of Pauli terms, and illustrated its improvement of success probability in simulations. Given the ability of efficiently implementing imaginary time evolution in quantum computers, one may explore the techniques for preparing thermal states or studying finite temperature properties of quantum systems. Furthermore, its application in non-Hermitian physics is also an expected prospect. By decomposing a non-Hermitian Hamiltonian into real and imaginary part, we can apply the ITE algorithm into the evolution of the imaginary part, which enables us to simulate the dynamical process of a generic non-Hermitian Hamiltonian.

\section{Acknowledgements}
This research was supported by National Basic Research Program of China. S.W. acknowledge the National Natural Science Foundation of China under Grants No. 12005015. We gratefully acknowledge support from the National Natural Science Foundation of China under Grants No. 11974205. The National Key Research and  Development Program of China (2017YFA0303700); The Key Research and  Development Program of Guangdong province (2018B030325002); Beijing Advanced Innovation Center for Future Chip (ICFC).
\\ 
\noindent

\appendix

\section{About Quafu}
\label{app:Quafu}

Quafu is an open cloud platform for quantum computation \cite{ref:quafu-web}. It provides four specifications of superconducting quantum processors currently, three of them support  general quantum logical gates, which are 10-qubits and 18-qubits processors with one-dimensional chain structure named P-10 and P-18, an $50+$ qubits processor with 2-dimensional honeycomb structure named P-50. 

\begin{figure}[htbp]
\includegraphics[width=7cm]{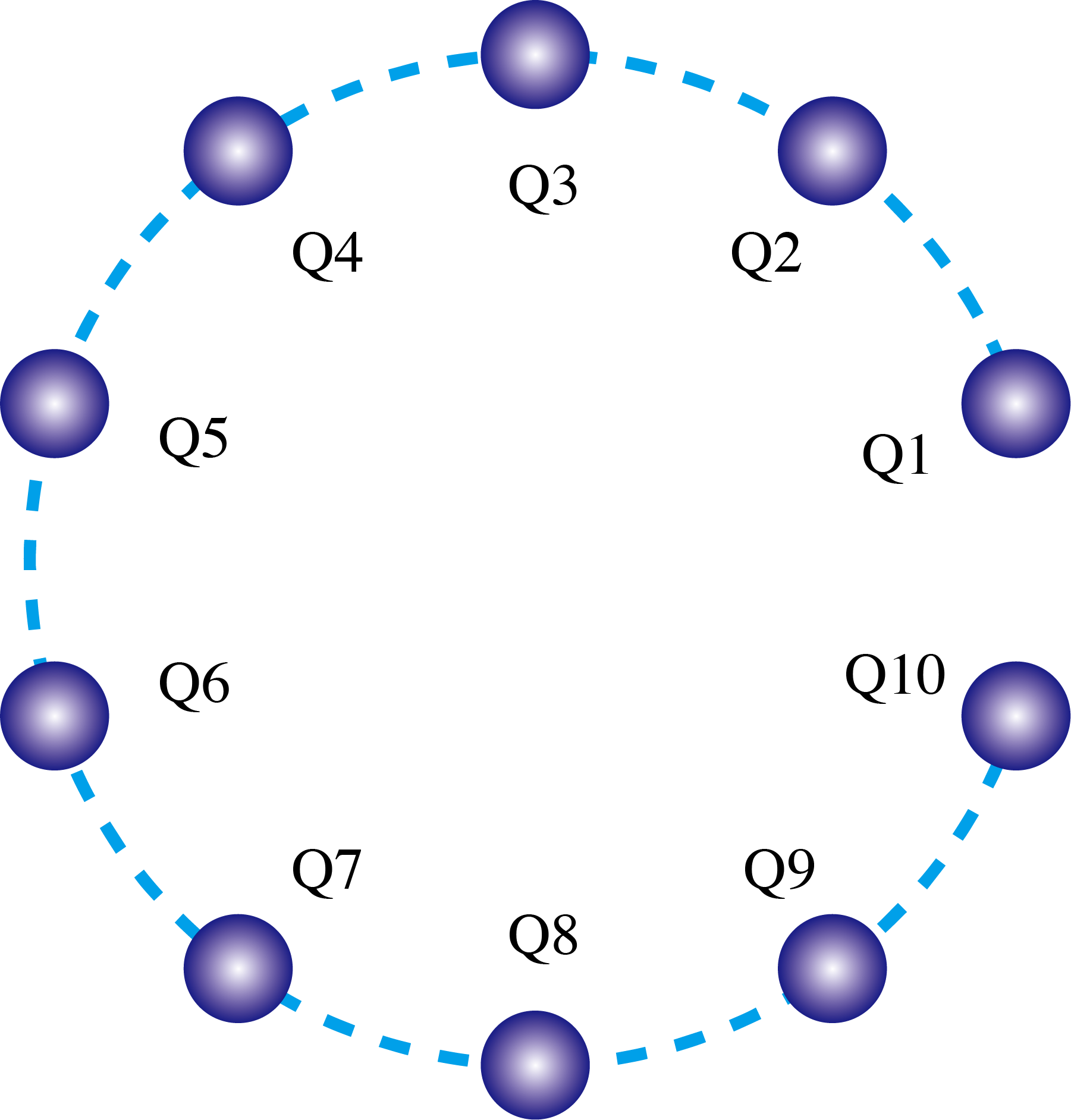}
\caption{The topological structure of quantum processor P-10. Each qubit capacitively coupled to its nearest-neighbors.
}
\label{fig:p10}
\end{figure}

PyQuafu is an open-source SDK for Python based on Quafu cloud platform. Users can easily install through pypi or source install with GitHub \cite{ref:pyquafu-github}. In this article, we use quantum processor of P-10 which is shown in Fig. \ref{fig:p10}. The processor consists of 10 transmon qubits ($Q_1-Q_{10}$) arrayed in a row, with each qubit capacitively coupled to its nearest-neighbors. Each transmon qubit can be modulated in frequency from about 4 to 5.7 GHz and excited to the excited state individually. All qubits can be probed though a common transmission line connected to their own readout resonators.The qubit parameters and coherence performance can be found in Table \ref{table:P10}. The idle frequencies of each qubit $\omega^{10}_j$ are designed to reduce residual coupling strength from other qubits. 

\begin{table*}
\caption{\label{table:P10}Device parameters. $\omega^{s}_{j}$ shows the maximum frequency of $Q_{j}$. $\omega^{10}_{j}$ corresponds to the idle frequency of $Q_{j}$. $\omega^{r}_{j}$ shows the resonant frequency of $Q_{j}$ during readout. $\eta_j$ corresponds to  the anharmonicity of $Q_{j}$. $g_{j,j+1}$ is the coupling strength between nearest-neighbor qubits. $T_{1,j}$ and $T^{*}_{2,j}$ represent the relaxation time and coherence time of $Q_{j}$. $F_{0,j}$ and $F_{1,j}$ are readout fidelities of $Q_{j}$ in $|0\rangle$ and $|1\rangle$. $F_{j,j+1}$ represents the fidelity of CZ gate composed of $Q_{i}$ and $Q_{j}$, which is obtained by randomized benchmarking.}
    \centering
    \setlength{\tabcolsep}{1.8mm}{
    \begin{tabular}{|ccccccccccc|} 
        \hline qubit&$Q_{1}$&$Q_{2}$&$Q_{3}$&$Q_{4}$&$Q_{5}$&$Q_{6}$&$Q_{7}$&$Q_{8}$&$Q_{9}$&$Q_{10}$\\
        \hline $\omega^{s}_{j}/2\pi$ (GHz)  &5.536 &5.069 &5.660 &4.742 &5.528 &4.929 &5.451 &4.920 &5.540 &4.960 \\
        $\omega^{10}_{j}/2\pi$  (GHz)    &5.456 &4.424 &5.606 &4.327 &5.473 &4.412 &5.392 &4.319 &5.490 &4.442 \\
        $\omega^{r}_{j}/2\pi$  (GHz) & 5.088 &4.702 &5.606 &4.466 &5.300 &4.804 &5.177 &4.697 &5.474 &4.819 \\
        $\eta_{j}/2\pi$   (GHz) & 0.250 &0.207 &0.251 &0.206 &0.251 &0.203 &0.252 &0.204 &0.246 &0.208 \\
        $g_{j,j+1}/2\pi$ (MHz) & 12.07 &11.58 &10.92 &10.84 &11.56 &10.00 &11.74 &11.70 &11.69 & - \\
        $T_{1,j}$ (us)    &20.0 &52.5 &15.9 &16.3 &36.9 &44.4 &30.8 &77.7 &22.8 &25.0 \\
        $T_{2,j}^{*}$ (us) &8.60 &1.48 &9.11 &2.10 &12.8 &2.73 &15.7 &1.88 &4.49 &2.05 \\
        $F_{0,j}$ (\%) & 98.90 &98.32 &98.67 &95.30 &97.00 &95.47 &97.00 &96.37 &98.33 &97.13 \\
        $F_{1,j}$ (\%) & 92.90 &92.30 &92.97 &91.53 &86.17 &87.93 &93.40 &93.37 &94.63 &92.07 \\
        $F_{j,j+1}$ (\%) &94.2 &97.8 &96.6 &97.3 &96.8 &97.0 &94.5 &93.2 &96.0 & -\\
        \hline
        
    \end{tabular}}
\end{table*}

\section{Unitary Transformation of Pauli Product Terms}
\label{app:Uk}

To realize $\widetilde{T}_k=e^{-c_k h_k\Delta t}$ in quantum circuits, we apply a unitary transformation
\begin{equation}
\label{eq:app:UchU}
U_k c_k h_k U_k^\dagger=-|c_k|\sigma_z^{l_k},\quad l_k\in\{1,\dots,n\}
\end{equation}
which gives us
\begin{equation}
\widetilde{T}_k=U_k^\dagger \exp\left(-|c_k|\sigma_z^{l_k}\Delta t\right)U_k.
\end{equation}

We will first prove that the unitary gate $U_k$ which satisfies Eq. \ref{eq:app:UchU}, where $h_k=\sigma_{\alpha_1}^1\otimes\dots\otimes\sigma_{\alpha_n}^n$, can be constructed with $\mathcal{O}(n)$ CNOT and single qubit gates. In fact, there are many methods to construct such $U_k$. Here we will show one method which is to decompose $U_k$ into three unitaries: $U_k=V_3 V_2 V_1$.

First, we notice that
\begin{equation}
H\sigma_x H=\sigma_z,\quad HS^\dagger\sigma_y SH =\sigma_z
\end{equation}
where $H$ is the Hadamard gate and $S$ is the $\pi/4$ phase gate (i.e. $S=e^{-i\pi\sigma_z/4}$). Thus, we construct $V_1$ by applying $H$ and $HS^\dagger$ respectively on those qubits whose corresponding Pauli matrix is $\sigma_x$ and $\sigma_y$. Then we have
\begin{equation}
V_1 c_k\left(\sigma_{\alpha_1}^1\otimes\dots\otimes\sigma_{\alpha_n}^n\right)V_1^\dagger
=c_k\sigma_{\gamma_1}^1\otimes\dots\otimes\sigma_{\gamma_n}^n
\end{equation}
where $\gamma_j=0$ if $\alpha_j=0$, and $\gamma_j=z$ otherwise.

Next, it is noticed that
\begin{equation}
C_{i,j}\left(\sigma_z^i\otimes\sigma_z^j\right)C_{i,j}=I^i\otimes\sigma_z^j
\end{equation}
where $C_{i,j}$ represents the CNOT gate with the $i$-th qubit being the control qubit and the $j$-th qubit being the target. To construct $V_2$, we first choose an arbitrary $l$ with $\gamma_l=z$, then apply $C_{j,l}$ for all $j$ with $\gamma_j=z,j\neq l$. Thus we have
\begin{equation}
\begin{split}
&V_2 V_1 c_k\left(\sigma_{\alpha_1}^1\otimes\dots\otimes\sigma_{\alpha_n}^n\right)V_1^\dagger V_2^\dagger=I^{\otimes l-1}\otimes c_k\sigma_z^l \otimes I^{\otimes n-l}
\end{split}
\end{equation}

For the last step, we have
\begin{equation}
\sigma_x\sigma_z\sigma_x=-\sigma_z,
\end{equation}
therefore if $c_k>0$, $V_3=\sigma_x^l$; otherwise $V_3=I$.

By now we have successfully constructed the unitary gate $U_k$ which satisfied Eq. \ref{eq:app:UchU} by $U_k=V_3 V_2 V_1$, and the maximum number of CNOT and single qubit gates used in this procedure is $2n+(n-1)+1=3n$, where $n$ is the number of qubit.

After the action of $U_k$, the state of work qubits is $U_k\ket{\Phi}$. We write it as $U_k\ket{\Phi}=a_0\ket{\psi_0}+a_1\ket{\psi_1}$, where $\ket{\psi_0}$ and $\ket{\psi_1}$ are the projection of the work-qubit state on the subspace where the $l_k$-th qubit is $\ket{0}$ and $\ket{1}$, respectively. Therefore,
\begin{equation}
\label{eq:app:Phi}
\ket{\Phi}=a_0 U_k^\dagger \ket{\psi_0}+a_1 U_k^\dagger \ket{\psi_1}.
\end{equation} 

From Eq. \ref{eq:app:UchU}, we have
\begin{equation}
c_k h_k U_k^\dagger \ket{\psi_0}=-|c_k|U_k^\dagger \sigma_z^{l_k} \ket{\psi_0}=-|c_k|U_k^\dagger\ket{\psi_0}
\end{equation}
which indicates $U_k^\dagger \ket{\psi_0}$ is an eigenstate of $c_k h_k$ with eigenvalue $-|c_k|$. Similarly, we can show $U_k^\dagger \ket{\psi_1}$ is an eigenstate of $c_k h_k$ with eigenvalue $|c_k|$. Therefore, from Eq. \ref{eq:app:Phi} we can say that $a_0,a_1$ are the amplitude of the projection of $\ket{\Phi}$ on the ground-state subspace and excited-state subspace of $c_k h_k$, respectively.

\section{Mapping the H$_2$ and LiH Hamiltonian to Qubits}
\label{app:molecules}

A molecule is a many-body system composed of nuclei and electrons. Its Hamiltonian includes the kinetic energy of each particle and the Coulomb potential energy between any two of these particles, written as
\begin{equation}
\begin{split}
\mathcal{H}=&-\sum_i \frac{1}{2M_i}\nabla_{R_i}^2-\sum_i \frac{1}{2}\nabla_{r_i}^2-\sum_{i,j}\frac{Z_j}{\left|r_i-R_j\right|^2} \\
&+\sum_{i,j}\frac{Z_i Z_j}{\left|R_i-R_j\right|^2}+\sum_{i,j}\frac{1}{\left|r_i-r_j\right|^2}
\end{split}
\end{equation}
in atomic units, where $M_i,Z_i,R_i$ and $r_i$ are the masses, charges, positions of nuclei and the positions of electrons, respectively. We first apply the Born-Oppenheimer approximation, which assumes the nuclear coordinates to be parameters rather than variables. Then the Hamiltonian is projected onto a chosen set of orbitals. Here we choose the standard Gaussian STO-3G basis \cite{STO3G}, and rewrite the Hamiltonian in the second-quantized form:
\begin{equation}
\mathcal{H}=\sum_{ij}u_{ij}a_i^\dagger a_j+\sum_{ijkl}u_{ijkl}a_i^\dagger a_j^\dagger a_k a_l+\dots
\end{equation}
where $a_i^\dagger$ and $a_j$ are the creation and annihilation operators of particle in the $i$-th and $j$-th orbital, respectively, and the $\dots$ represents the high-order interactions.

To map the fermionic Hamiltonian to the qubit Hamiltonian, we use the Jordan-Wigner transformation (JWT), which could transform the creation and annihilation operators into Pauli matrices. Under the JWT, the state of the $j$-th qubit $\ket{0}$ or $\ket{1}$ respectively corresponds to the $j$-th orbital being unoccupied or occupied.

For H$_2$ molecules, we use the method described in Supplementary Material of Ref. \cite{Colless18} to get the qubit Hamiltonian with 2 qubits, which is written as
\begin{equation}
\label{eq:app:H2}
\mathcal{H}_{\text{H}_2}=c_0+c_1\sigma_z^1+c_1\sigma_z^2+c_2\sigma_z^1\sigma_z^2
+c_3\sigma_x^1\sigma_x^2,
\end{equation}
The exact coefficients used in our work are shown in Table \ref{table:H2}.

For LiH molecules, we assume perfect filling of the innermost two $1s$ spin orbitals of Li, and define the Hamiltonian on the basis of the $2s,\ 2p_x$ orbitals that are associated with Li and the $1s$ orbitals that are associated with H, for a total of 6 spin orbitals. After the JWT, we obtain a 6-qubit Hamiltonian. We explicitly list the LiH Hamiltonian at the bound distance in Table \ref{table:LiH}.

\section{Experiments to Simulate H$_2$ on the Superconducting and Trapped Ion QPU}
\label{app:exp}

\begin{figure}
\includegraphics[width=8.5cm]{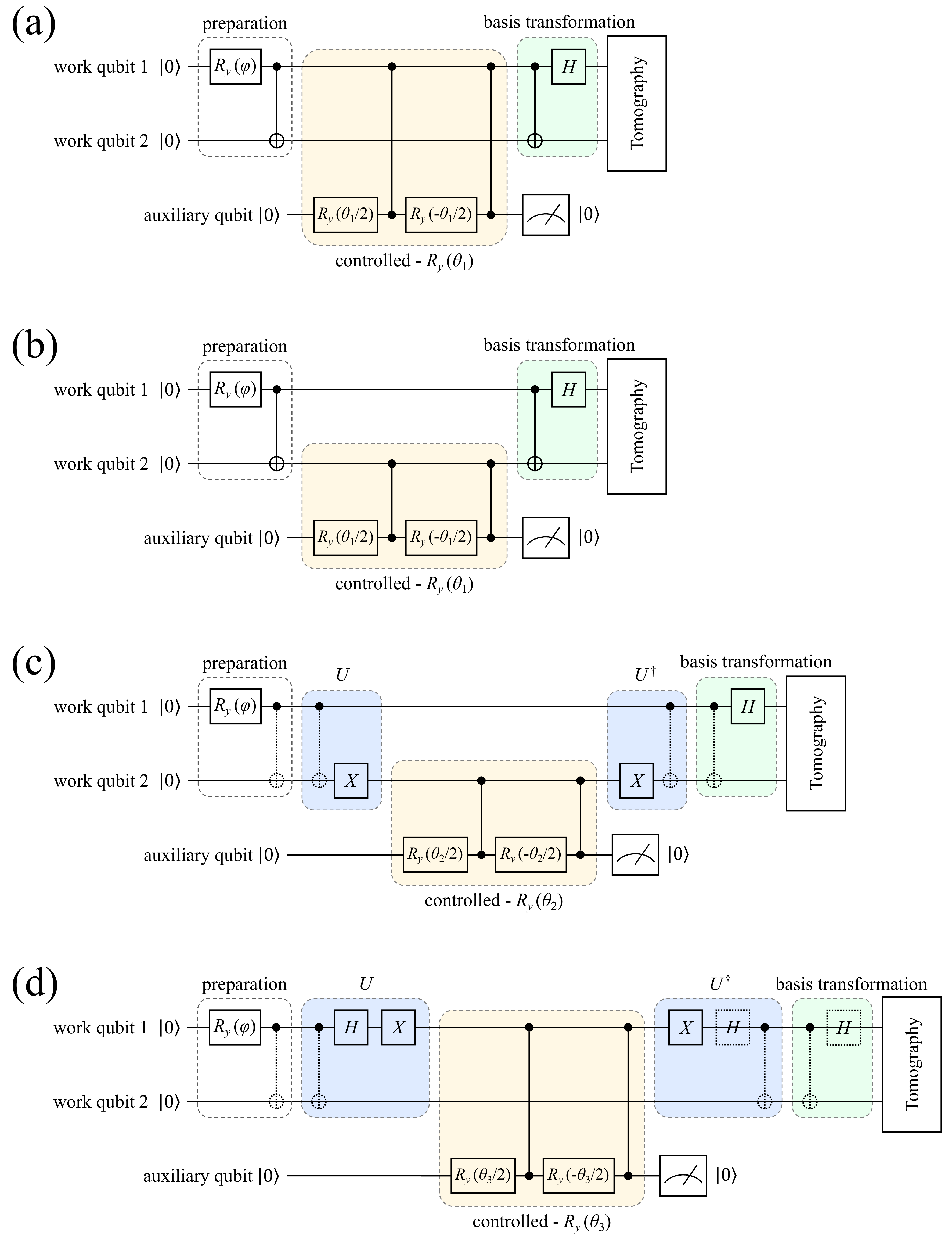}
\caption{Quantum circuits for applying PITE on H$_2$ molecules in our experiments. (a)-(d) correspond to the non-unitary evolution operators $\widetilde{T}_k\ (k=1,2,3,4)$, respectively. The quantum gates with dashed lines cancel out with one another, so they are not carried out in the experiments.}
\label{circuit:H2}
\end{figure}

The two-qubit H$_2$ Hamiltonian (Eq. \ref{eq:app:H2}) contains 4 non-identity terms, each corresponding to a non-unitary evolution operator when we apply the PITE,
\begin{equation}
\begin{split}
\label{eq:app:Tk}
&\widetilde{T}_1=\exp\left(-c_1\sigma_z^1\Delta t\right), \\
&\widetilde{T}_2=\exp\left(-c_1\sigma_z^2\Delta t\right), \\
&\widetilde{T}_3=\exp\left(-c_2\sigma_z^1\sigma_z^2\Delta t\right), \\
&\widetilde{T}_4=\exp\left(-c_3\sigma_x^1\sigma_x^2\Delta t\right).
\end{split}
\end{equation}
In the experiments, we need to apply $\widetilde{T}_1,\widetilde{T}_2,\widetilde{T}_3,\widetilde{T}_4$ on the work qubits in turn, as a cycle. And in the end of each cycle, we need to measure the energy expectation value to show its convergence. We use the 4 quantum circuits shown in Fig. \ref{circuit:H2}(a)-(d) to implement $\widetilde{T}_k\ (k=1,2,3,4)$, respectively. These circuits are composed of following parts:

(1) $U$ and $U^\dagger$ (blue blocks) correspond to $U_k$ and $U_k^\dagger$ in Fig. \ref{circuit:Tk}.

(2) Controlled-$R_y$ (orange blocks) corresponds to the controlled-$R_y$ gate in Fig. \ref{circuit:Tk}.

(3) Basis transformation (green blocks): If we directly measure the work qubits on ZZ basis after $U^\dagger$, the theoretical probability of obtaining 11 is on the order of $10^{-3}\sim10^{-2}$, which is easily affected by the measurement error. To reduce this influence, we measure the work qubits on XX basis by applying the basis transformation before the measurement.

(4) Preparation and Tomography (white blocks): As the end of a Trotter step, the quantum state of work qubits after basis transformation is tomographed. Then we calculate the proper state numerically by reversing the basis transformation, and prepare the state in the next quantum circuit as the input state. As the state only evolves in the real-coefficient subspace spanned by $\ket{00}$ and $\ket{11}$, the preparation only requires a $R_y$ gate and a CNOT gate. The angular parameter $\phi$ of $R_y$ can be obtained from the tomography result. In the experiments, each quantum state is tomographed for three times (with 2000 shots on superconducting QPU and 1000 shots on trapped ion QPU for each time). We calculate $\phi$ for each time, and take their average value as the input parameter of the next Trotter step. For the first Trotter step, the input state is simply $\ket{\Phi_\text{HF}}=\ket{00}$, i.e. the input parameter $\phi=0$. Besides, after the tomography of the $\widetilde{T}_4$ circuit, we also use the proper state to calculate the energy value $\braket{E}=\braket{\Phi|\mathcal{H}_{\text{H}_2}|\Phi}$.

\section{Derivation of Output Fidelity and the Lower Bound of success Probability}
\label{app:complexity}

A quantum state $\ket{\Phi_0}$ is the linear combination of all eigenstates of the Hamiltonian:
\begin{equation}
\ket{\Phi_0}=\sum_{i=0}\mu_i\ket{E_i}
\end{equation}
Suppose the eigenvalues $E_i$ are ordered as $E_0\leq E_1\leq E_2\leq\dots$. After the evolution under the imaginary time $\beta$, the fidelity between $\ket{\Phi_\beta}=Ae^{-\beta\mathcal{H}}\ket{\Phi_0}$ ($A$ is the normalization factor) and the exact ground state $\ket{\psi_G}=\ket{E_0}$ is
\begin{equation}
\begin{split}
\mathcal{F}(\ket{\Phi_\beta},\ket{\psi_G})&=\frac{\braket{\Phi_0|e^{-\beta\mathcal{H}}|E_0}\braket{E_0|e^{-\beta\mathcal{H}}|\Phi_0}}{\braket{\Phi_0|e^{-2\beta\mathcal{H}}|\Phi_0}} \\
&=\frac{|\mu_0|^2 e^{-2\beta E_0}}{\sum_{i=0}|\mu_i|^2 e^{-2\beta E_i}} \\
&=\frac{s_0}{s_0+\sum_{i=1}s_i e^{-2\beta\Omega_i}}
\end{split}
\end{equation}
where $s_i=|\mu_i|^2$ and $\Omega_i=E_i-E_0$. To take a lower bound of $\mathcal{F}$, we notice that $\Omega_i\geq\Omega_1$ for all $i\geq 1$. Therefore, with $\sum_{i=0} s_i=1$, we have
\begin{equation}
\mathcal{F}\geq\frac{s_0}{s_0+\sum_{i=1}s_i e^{-2\beta\Omega_1}}=\frac{s_0}{s_0+(1-s_0)e^{-2\beta\Omega_1}}
\end{equation}
where $s_0=|\braket{E_0|\Phi_0}|^2$ is the initial fidelity. This is the proof for Eq. \ref{eq:fidelity_beta}. We can find the relation between the imaginary time length $\beta$ and the output fidelity error $\epsilon=1-\mathcal{F}$:
\begin{equation}
\label{eq:app:beta.omega}
\beta\Omega_1\leq\frac{1}{2}\ln\left(\frac{1-s_0}{s_0}\frac{1-\epsilon}{\epsilon}\right)
\end{equation}
Here we write the product of $\beta$ and $\Omega_1$ because we can choose the scale of the Hamiltonian to change the value of $\Omega_1$, and $\beta$ is also dependent on the scale of the Hamiltonian. So it is the product $\beta\Omega_1$ that really matters.

Now we look into the lower bound of the success probability of the PITE algorithm. The simulation results show that far higher success probability can be reached than be suggested by the RLB (Eq. \ref{eq:prob_f_rlb}). We now give a more practical estimation of success probability. We will do the derivation for the generalized PITE algorithm, and degenerate the conclusion for the original PITE.

In the generalized PITE algorithm, the Hamiltonian is $\mathcal{H}=\sum_k H[k]$, and $\lambda[k]_i,\ket{\lambda[k]_i}$ are the eigenvalues and eigenvectors of $H[k]$. The probability of successfully implementing $e^{-H[k]\Delta t}$ is
\begin{equation}
\label{eq:app:P_k_dt}
P[k](\Delta t)=|a_0|^2+\sum_{i=1}|a_i|^2 e^{-2\omega[k]_i\Delta t}
\end{equation}
where $a_i=\braket{\lambda[k]_i|\Phi}$ ($i\geq 0$), $\omega[k]_i=\lambda[k]_i-\lambda[k]_0$ ($\lambda[k]_0$ is the lowest eigenvalue of $H[k]$). Take the approximation of Eq. \ref{eq:app:P_k_dt} as
\begin{equation}
\begin{split}
P[k](\Delta t)&\approx|a_0|^2+\sum_{i=1}|a_i|^2\left(1-2\omega[k]_i\Delta t\right) \\
&=1-2\Delta t\sum_{i=1}|a_i|^2\omega[k]_i \\
&=1-2\Delta t\left(\braket{\Phi|H[k]|\Phi}-\lambda[k]_0\right) \\
&\approx\exp\left[-2\Delta t\left(\braket{\Phi|H[k]|\Phi}-\lambda[k]_0\right)\right].
\end{split}
\end{equation}
Then the total success probability after $\Delta t$ is approximated by
\begin{equation}
P(\Delta t)\approx\exp\left[-2\Delta t\left(\braket{\Phi|\mathcal{H}|\Phi}-\sum_k\lambda[k]_0\right)\right]
\end{equation}
Here we are approximating that $\ket{\Phi}$ is constant during one evolution step. Thus the final success probability after imaginary time $\beta$ is
\begin{equation}
P_\text{final}\approx\exp\left[-2\int_0^\beta\mathrm{d}t\left(\braket{\Phi|\mathcal{H}|\Phi}-\sum_k\lambda[k]_0\right)\right]
\end{equation}

To do the integral in this equation, we need
\begin{equation}
\begin{split}
\braket{\Phi_t|\mathcal{H}|\Phi_t}&=\frac{\braket{\Phi_0|e^{-\mathcal{H}t}H e^{-\mathcal{H}t}|\Phi_0}}{\braket{\Phi_0|e^{-2\mathcal{H}t}|\Phi_0}} \\
&=\frac{\sum_{i=0}|\mu_i|^2 E_i e^{-2E_i t}}{\sum_{i=0}|\mu_i|^2 e^{-2E_i t}} \\
&=E_0+\frac{\sum_{i=1}s_i \Omega_i e^{-2\Omega_i t}}{s_0+\sum_{i=1}s_i e^{-2\Omega_i t}} \\
&\leq E_0+\frac{(1-s_0)\Omega_\text{max} e^{-2\Omega_1 t}}{s_0}
\end{split}
\end{equation}
where $\Omega_\text{max}$ is the maximum of all $\Omega_i$, namely the gap between the highest excited state and the ground state of the Hamiltonian. Therefore,
\begin{equation}
\begin{split}
\int_0^\beta\mathrm{d}t\braket{\Phi|\mathcal{H}|\Phi}&\leq\beta E_0+\frac{(1-s_0)\Omega_\text{max}}{s_0}\int_0^\beta\mathrm{d}t\ e^{-2\Omega_1 t} \\
&=\beta E_0+\frac{(1-s_0)\Omega_\text{max}}{2s_0\Omega_1}\left(1-e^{-2\beta\Omega_1}\right).
\end{split}
\end{equation}
Thus the lower bound (ALB) of the final success probability is
\begin{equation}
\label{eq:app:prob_f_alb}
\begin{split}
&P_\text{final}\geq P_\text{ALB} \\
&\sim\exp\left[-2\beta\left(E_0-\sum_k\lambda[k]_0\right)
-\frac{(1-s_0)\Omega_\text{max}}{s_0\Omega_1}\left(1-e^{-2\beta\Omega_1}\right)\right].
\end{split}
\end{equation}
Especially, when $H[k]=c_k h_k$ are Pauli terms, $\lambda[k]_0=-|c_k|$, which turns Eq. \ref{eq:app:prob_f_alb} into Eq. \ref{eq:prob_f_alb}. We can see that when $\mathcal{H}$ is given, the only changeable part in Eq. \ref{eq:app:prob_f_alb} is $\sum_k\lambda[k]_0$. By increasing this part we can enlargement the success probability.

Furthermore, substituting Eq. \ref{eq:app:beta.omega} into Eq. \ref{eq:app:prob_f_alb}, we obtain the relation between the success probability and the output error:
\begin{equation}
\begin{split}
P_\text{ALB}&\geq\left(\frac{s_0}{1-s_0}\frac{\epsilon}{1-\epsilon}\right)^\kappa\exp\left[-\frac{\Omega_\text{max}}{s_0\Omega_1}\left(1-\frac{s_0}{1-\epsilon}\right)\right] \\
&=\mathcal{O}\left(\left(\frac{\epsilon}{1-\epsilon}\right)^\kappa\right)
\end{split}
\end{equation}
where $\kappa=\left(E_0-\sum_k\lambda[k]_0\right)/\Omega_1$ is independent on the scale of Hamiltonian.

\begin{figure*}[h]
\includegraphics[width=14cm]{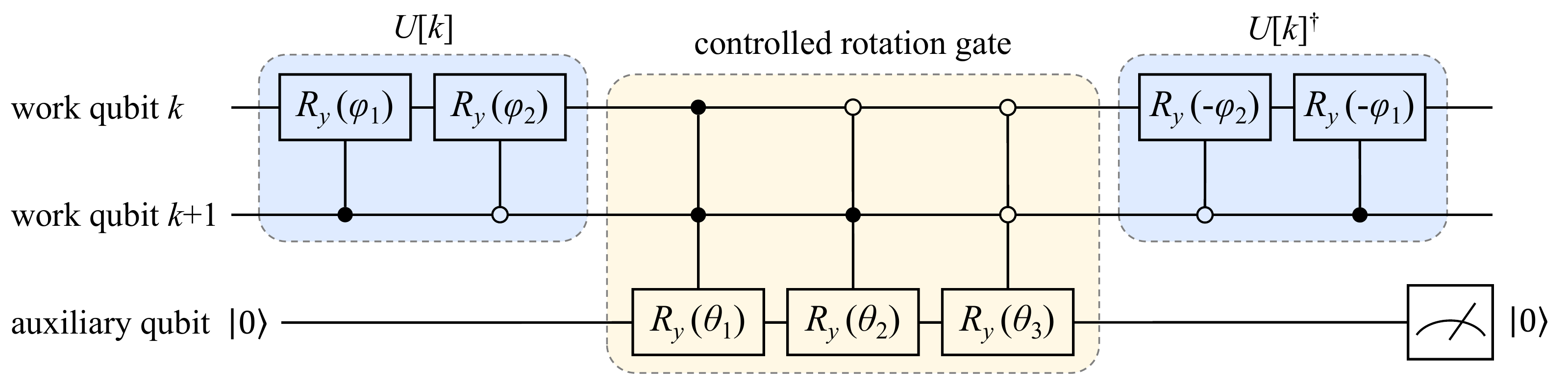}
\caption{Quantum circuit that implements $e^{-H[k]\Delta t}$ for solving the quantum Ising model.}
\label{circuit:Ising}
\end{figure*}

\section{Simulation using Generalized PITE}
\label{app:generalized}

For the quantum Ising cyclic chain, we rewrite its Hamiltonian as $\mathcal{H}=\sum_{k=1}^n H[k]$ with
\begin{equation}
H[k]=-\left(\sigma_z^k \sigma_z^{k+1}+g\sigma_x^k+h\sigma_z^k\right)
\end{equation}
in the case where $J=1$. Thus every $H[k]$ is a local operator acting on the $k$-th and $(k+1)$-th qubit. We can write the matrix form of $H[k]$ in the two-qubit computational basis:
\begin{equation}
H[k]=-\begin{pmatrix} 1+h & 0 & g & 0 \\ 0 & -1+h & 0 & g \\
g & 0 & -1-h & 0 \\ 0 & g & 0 & 1-h \end{pmatrix}.
\end{equation}
Its eigenvalues are $\lambda_{0,3}=\mp\sqrt{g^2+(h+1)^2}$, $\lambda_{1,2}=\mp\sqrt{g^2+(h-1)^2}$, and the corresponding eigenstates $\ket{\lambda_{0,1,2,3}}$ are also known. Following the procedure given in \ref{sec:generalized}, we can use the quantum circuit shown in Fig \ref{circuit:Ising} to implement $e^{-H[k]\Delta t}$, and the angles of the rotation gates are
\begin{equation}
\begin{split}
&\phi_1=\cos^{-1}\frac{1-h}{\sqrt{g^2+(h-1)^2}} \\
&\phi_2=\cos^{-1}\frac{-1-h}{\sqrt{g^2+(h+1)^2}} \\
&\theta_i=2\cos^{-1}(e^{-\Delta t(\lambda_i-\lambda_0)}),\ i=1,2,3.
\end{split}
\end{equation}

For the LiH molecule, the Hamiltonian has 62 Pauli terms (including the identity term). After neglecting the identity, we rearrange the other 61 terms and group them into 22 sets, with each set corresponding a $H[k']$ used in the generalized PITE algorithm. When doing this, we follow guidance of increasing $\sum_k\lambda[k]_0$ (see details in Appendix \ref{app:complexity}) and the principle that the eigen-systems of every $H[k]$ should be easily known through classical computation. The grouping of the Pauli terms in LiH Hamiltonian at its bound distance is shown in Table \ref{table:LiH}, and the same grouping strategy is employed at other interatomic distances.

\section{Simulating Noises using Quantum Channel}
\label{app:noise}

The quantum noise is described by quantum channel:
\begin{equation}
\mathcal{E}(\rho)=\sum_{\nu} \hat{E}_\nu \rho\hat{E}_\nu^\dagger
\end{equation}
where $\rho$ is the density matrix of the system and $\hat{E}_\nu$ are Kraus operators, satisfying
\begin{equation}
\sum_{\nu} \hat{E}_\nu^\dagger\hat{E}_\nu=I.
\end{equation}
For a single qubit in a quantum circuit, the main sources of quantum noise are qubit relaxation and dephasing, which correspond to the three Kraus operators shown in Eq. \ref{eq:Kraus}. For $n$-qubit cases, the number of Kraus operator is $3^n$, and each Kraus operator can be described by $\hat{E}_{\nu'}\in\{\hat{E}_1,\hat{E}_2,\hat{E}_3\}^{\otimes n}$.

\begin{table*}
\caption{\label{table:H2}The coefficients in H$_2$ Hamiltonian at different interatomic distance $R$.}
\renewcommand{\arraystretch}{1.2}
\begin{tabular}{|m{1cm}<{\centering}|m{2cm}<{\centering}|m{2cm}<{\centering}|m{2cm}<{\centering}|m{2cm}<{\centering}|}
\hline
$R$(\AA) & $c_0$ & $c_1$ & $c_2$ & $c_3$ \\
\hline \hline
0.35 & 7.01273E-01 & -7.47416E-01 & 1.31036E-02 & 1.62573E-01 \\
\hline
0.45 & 2.67547E-01 & -6.33890E-01 & 1.27192E-02 & 1.66621E-01 \\
\hline
0.55 & -1.83734E-02 & -5.36489E-01 & 1.23003E-02 & 1.71244E-01 \\
\hline
0.65 & -2.13932E-01 & -4.55433E-01 & 1.18019E-02 & 1.76318E-01 \\
\hline
0.75 & -3.49833E-01 & -3.88748E-01 & 1.11772E-02 & 1.81771E-01 \\
\hline
0.85 & -4.45424E-01 & -3.33747E-01 & 1.04061E-02 & 1.87562E-01 \\
\hline
1.05 & -5.62600E-01 & -2.48783E-01 & 8.50998E-03 & 1.99984E-01 \\
\hline
1.25 & -6.23223E-01 & -1.86173E-01 & 6.45563E-03 & 2.13102E-01 \\
\hline
1.45 & -6.52661E-01 & -1.38977E-01 & 4.59760E-03 & 2.26294E-01 \\
\hline
\end{tabular}
\end{table*}

\begin{table*}
\caption{\label{table:LiH}The LiH Hamiltonian at bound distance, as well as the grouping of the Pauli terms used in generalized PITE.}
\renewcommand{\arraystretch}{1.35}
\begin{tabular}{|m{0.8cm}<{\centering}|m{2.2cm}<{\centering}|m{2cm}<{\centering}|m{1cm}<{\centering}||m{0.8cm}<{\centering}|m{2.2cm}<{\centering}|m{2.8cm}<{\centering}|m{1cm}<{\centering}|}
\hline
$k$ & $c_k$ & $h_k$ & $H[k']$ & $k$ & $c_k$ & $h_k$ & $H[k']$ \\
\hline \hline
1 & -7.35094E+00 & $\bm{1}$ & ------ & 33 & -1.49854E-03 & $\sigma_x^2 \sigma_z^3 \sigma_x^4 \sigma_z^5$ & \multirow{2}{*}{$H[13]$} \\
\cline{1-7}
2 & -1.58950E-01 & $\sigma_z^1$ & \multirow{3}{*}{$H[1]$} & 34 & -1.49854E-02 & $\sigma_y^2 \sigma_z^3 \sigma_y^4 \sigma_z^5$ & \\
\cline{1-3} \cline{5-8}
3 & -1.58950E-01 & $\sigma_z^2$ & & 35 & 1.13678E-02 & $\sigma_x^1 \sigma_z^2 \sigma_x^3 \sigma_z^6$ & \multirow{2}{*}{$H[14]$} \\
\cline{1-3} \cline{5-7}
4 & 7.82811E-02 & $\sigma_z^1 \sigma_z^2$ & & 36 & 1.13678E-02 & $\sigma_y^1 \sigma_z^2 \sigma_y^3 \sigma_z^6$ & \\
\hline
5 & -1.45795E-01 & $\sigma_z^3$ & \multirow{3}{*}{$H[2]$} & 37 & -1.17598E-03 & $\sigma_z^1 \sigma_x^2 \sigma_z^3 \sigma_x^4$ & \multirow{2}{*}{$H[15]$} \\
\cline{1-3} \cline{5-7}
6 & -1.45795E-01 & $\sigma_z^4$ & & 38 & -1.17598E-03 & $\sigma_z^1 \sigma_y^2 \sigma_z^3 \sigma_y^4$ & \\
\cline{1-3} \cline{5-8}
7 & 8.51132E-02 & $\sigma_z^3 \sigma_z^4$ & & 39 & 3.56300E-03 & $\sigma_x^1 \sigma_z^2 \sigma_x^3 \sigma_z^5$ & \multirow{2}{*}{$H[16]$} \\
\cline{1-7}
8 & 2.96723E-02 & $\sigma_z^5$ & \multirow{3}{*}{$H[3]$} & 40 & 3.56300E-03 & $\sigma_y^1 \sigma_z^2 \sigma_y^3 \sigma_z^5$ & \\
\cline{1-3} \cline{5-8}
9 & 2.96723E-02 & $\sigma_z^6$ & & 41 & 3.56300E-03 & $\sigma_x^2 \sigma_z^3 \sigma_x^4 \sigma_z^6$ & \multirow{2}{*}{$H[17]$} \\
\cline{1-3} \cline{5-7}
10 & 1.24302E-01 & $\sigma_z^5 \sigma_z^6$ & & 42 & 3.56300E-03 & $\sigma_y^2 \sigma_z^3 \sigma_y^4 \sigma_z^6$ & \\
\hline
11 & 5.36162E-02 & $\sigma_z^1 \sigma_z^3$ & \multirow{3}{*}{$H[4]$} & 43 & -1.03458E-02 & $\sigma_x^1 \sigma_x^2 \sigma_y^3 \sigma_y^4$ & \multirow{4}{*}{$H[18]$} \\
\cline{1-3} \cline{5-7}
12 & 6.03396E-02 & $\sigma_z^3 \sigma_z^5$ & & 44 & -1.03458E-02 & $\sigma_y^1 \sigma_y^2 \sigma_x^3 \sigma_x^4$ & \\
\cline{1-3} \cline{5-7}
13 & 6.28713E-02 & $\sigma_z^1 \sigma_z^5$ & & 45 & 1.03458E-02 & $\sigma_x^1 \sigma_y^2 \sigma_y^3 \sigma_x^4$ & \\
\cline{1-7}
14 & 5.64568E-02 & $\sigma_z^1 \sigma_z^4$ & \multirow{3}{*}{$H[5]$} & 46 & 1.03458E-02 & $\sigma_y^1 \sigma_x^2 \sigma_x^3 \sigma_y^4$ & \\
\cline{1-3} \cline{5-8}
15 & 6.03396E-02 & $\sigma_z^4 \sigma_z^6$ & & 47 & -2.84063E-03 & $\sigma_x^3 \sigma_x^4 \sigma_y^5 \sigma_y^6$ & \multirow{4}{*}{$H[19]$} \\
\cline{1-3} \cline{5-7}
16 & 6.87743E-02 & $\sigma_z^1 \sigma_z^6$ & & 48 & -2.84063E-03 & $\sigma_y^3 \sigma_y^4 \sigma_x^5 \sigma_x^6$ & \\
\cline{1-4} \cline{5-7}
17 & 5.36162E-02 & $\sigma_z^2 \sigma_z^4$ & \multirow{3}{*}{$H[6]$} & 49 & 2.84063E-03 & $\sigma_x^3 \sigma_y^4 \sigma_y^5 \sigma_x^6$ & \\
\cline{1-3} \cline{5-7}
18 & 6.87743E-02 & $\sigma_z^2 \sigma_z^5$ & & 50 & 2.84063E-03 & $\sigma_y^3 \sigma_x^4 \sigma_x^5 \sigma_y^6$ & \\
\cline{1-3} \cline{5-8}
19 & 7.06853E-02 & $\sigma_z^4 \sigma_z^5$ & & 51 & -5.90301E-03 & $\sigma_x^1 \sigma_x^2 \sigma_y^5 \sigma_y^6$ & \multirow{4}{*}{$H[20]$} \\
\cline{1-4} \cline{5-7}
20 & 5.64568E-02 & $\sigma_z^2 \sigma_z^3$ & \multirow{3}{*}{$H[7]$} & 52 & -5.90301E-03 & $\sigma_y^1 \sigma_y^2 \sigma_x^5 \sigma_x^6$ & \\
\cline{1-3} \cline{5-7}
21 & 6.28713E-02 & $\sigma_z^2 \sigma_z^6$ & & 53 & 5.90301E-03 & $\sigma_x^1 \sigma_y^2 \sigma_y^5 \sigma_x^6$ & \\
\cline{1-3} \cline{5-7}
22 & 7.06853E-02 & $\sigma_z^3 \sigma_z^6$ & & 54 & 5.90301E-03 & $\sigma_y^1 \sigma_x^2 \sigma_x^5 \sigma_y^6$ & \\
\hline
23 & -1.49854E-03 & $\sigma_x^1 \sigma_x^3$ & \multirow{2}{*}{$H[8]$} & 55 & -4.73898E-03 & $\sigma_x^2 \sigma_x^3 \sigma_x^5 \sigma_x^6$ & \multirow{4}{*}{$H[21]$} \\
\cline{1-3} \cline{5-7}
24 & -1.49854E-03 & $\sigma_y^1 \sigma_y^3$ & & 56 & -4.73898E-03 & $\sigma_y^2 \sigma_y^3 \sigma_y^5 \sigma_y^6$ & \\
\cline{1-7}
25 & 1.13678E-02 & $\sigma_x^2 \sigma_x^4$ & \multirow{2}{*}{$H[9]$} & 57 & -4.73898E-03 & $\sigma_x^2 \sigma_y^3 \sigma_y^5 \sigma_x^6$ & \\
\cline{1-3} \cline{5-7}
26 & 1.13678E-02 & $\sigma_y^2 \sigma_y^4$ & & 58 & -4.73898E-03 & $\sigma_y^2 \sigma_x^3 \sigma_x^5 \sigma_y^6$ & \\
\hline
27 & 1.04793E-02 & $\sigma_x^1 \sigma_z^2 \sigma_x^3$ & \multirow{2}{*}{$H[10]$} & 59 & -4.73898E-03 & $\sigma_x^1 \sigma_z^2 \sigma_z^3 \sigma_x^4 \sigma_y^5 \sigma_y^6$ & \multirow{4}{*}{$H[22]$} \\
\cline{1-3} \cline{5-7}
28 & 1.04793E-02 & $\sigma_y^1 \sigma_z^2 \sigma_y^3$ & & 60 & -4.73898E-03 & $\sigma_y^1 \sigma_z^2 \sigma_z^3 \sigma_y^4 \sigma_x^5 \sigma_x^6$ & \\
\cline{1-7}
29 & 1.04793E-02 & $\sigma_x^2 \sigma_z^3 \sigma_x^4$ & \multirow{2}{*}{$H[11]$} & 61 & 4.73898E-03 & $\sigma_x^1 \sigma_z^2 \sigma_z^3 \sigma_y^4 \sigma_y^5 \sigma_x^6$ & \\
\cline{1-3} \cline{5-7}
30 & 1.04793E-02 & $\sigma_y^2 \sigma_z^3 \sigma_y^4$ & & 62 & 4.73898E-03 & $\sigma_y^1 \sigma_z^2 \sigma_z^3 \sigma_x^4 \sigma_x^5 \sigma_y^6$ & \\
\hline
31 & -1.17598E-03 & $\sigma_x^1 \sigma_z^2 \sigma_x^3 \sigma_z^4$ & \multirow{2}{*}{$H[12]$} & \multicolumn{4}{c}{\multirow{2}{*}{}} \\
\cline{1-3}
32 & -1.17598E-03 & $\sigma_y^1 \sigma_z^2 \sigma_y^3 \sigma_z^4$ & & \multicolumn{4}{c}{} \\
\cline{1-4}
\end{tabular}
\end{table*}



\bibliography{PITE.bib}

\end{document}